\documentclass{aa}
 
\usepackage{graphics} 
\usepackage{epsf} 
\begin{document}  
\thesaurus{08.16.4;
09.04.1;
10.03.1;
13.09.6;
} 
 
\title{Interstellar extinction towards the inner Galactic Bulge\thanks{Based 
on observations collected at the European Southern Observatory, La Silla Chile} }

\author{ 
M. Schultheis\inst{1} 
\and S.~Ganesh\inst{1,2}  
\and G.~Simon\inst{3}  
\and A.~Omont\inst{1}  
\and C.~Alard\inst{3,1}  
\and J.~Borsenberger\inst{1}  
\and E.~Copet\inst{4,1}  
\and N.~Epchtein\inst{6}  
\and P.~Fouqu\'e\inst{5}  
\and H.~Habing \inst{7}  
}  
\offprints{schulthe@iap.fr}

\institute{  
Institut d'Astrophysique de Paris, CNRS, 98bis Bd Arago, F-75014 Paris     
\and Physical Research Laboratory, Navarangpura, Ahmedabad 380009, India   
\and DASGAL CNRS UMR 8633, Observatoire de Paris, France                                 
\and DESPA  CNRS-UMR 8632,  Observatoire de Paris, France                                  
\and ESO, Santiago, Chile                                                  
\and O.C.A., Nice, France                                                  
\and Leiden Observatory, Leiden, The Netherlands                           
}  
  
\voffset 1.0true cm

\date{Received June 1999  / Accepted August 1999}  
  
\maketitle  
\sloppy

\begin{abstract}  
  
DENIS observations in the J (1.25\,$\mu$m) and $\rm K_{S}$ (2.15\,$\mu$m) bands together with  
isochrones calculated for the RGB and  
AGB phase are used to draw an extinction  map of the inner Galactic Bulge.  
The uncertainty in this method is mainly limited by the optical depth  
of the Bulge itself. A comparison with  fields of known extinction  
shows a very good agreement. We present an extinction map  
for the  inner Galactic Bulge ($\rm \sim 20\,deg^{2}$)


\end{abstract}

  
\section{Introduction} \label{introduction}  
Studying the stellar populations in the Galactic Bulge requires knowledge of  
the  interstellar extinction. Previous studies (e.~g. Catchpole et  
al. \cite{Catchpole90}, Frogel et al. \cite{Frogel99}, Unavane et al. \cite{Unavane98}) have shown that it can vary  
from $\rm A_{V} = 5^{m}$ up to $\rm  
A_{V} = 35^{m}$ towards the Galactic Centre region. As in most  parts of the  
Galactic Bulge, interstellar absorption is not homogeneous but occurs in  
clumps, a detailed  extinction map of the Bulge is therefore essential.  
  
 Catchpole et al.  
(\cite{Catchpole90}) mapped the interstellar extinction  around the Galactic Centre ($\rm \sim 2\,deg^{2}$)  
using the red giant branch of 47\,Tuc as a reference. Stanek et al. (\cite{Stanek96})  
 mapped the interstellar extinction of Baade's window using OGLE photometry  of red clump stars.  
Frogel et al. (\cite{Frogel99}) determined the interstellar reddening for a few fields in the  
inner Galactic Bulge using the red giant branch of Baade's window as a reference.  
$\rm A_{K}$ values ranging from 0.27 up to 2.15\,mag were found by the latter.  

The DENIS survey (Epchtein et al. \cite{Epchtein97}, Persi et al. \cite{Persi95}) with  2MASS (Skrutskie et 
al. \cite{Skrutskie97}) is the first attempt to carry out a complete survey  of the southern sky.  The 
limiting magnitudes in the three near-IR  photometric bands (I = 0.8\,$\mu$m, J = 1.25\,$\mu$m and $\rm K_{S} = 
2.15\, \mu$m) for point sources are $\rm 18^{m}$, $\rm 16^{m}$ and $\rm 13^{m}$ respectively.  The 
photometric accuracy (rms) is better than 0.1\,mag and the astrometric accuracy better than 1\,arcsec.   
These numbers are for uncrowded fields.  

DENIS $\rm K_{S}/(J-K_{S})$ colour magnitude diagrams (CMDs) in regions of low extinction show a well  
defined RGB and AGB sequence in the  Galactic Bulge (see Fig.~1). Due to their  high  
luminosities (up to $\rm M_{bol} \sim -4.5$ for non-saturated sources in DENIS) these  
stars are ideal tracers of the stellar populations in the Bulge and are found even in highly  
extincted regions. In this paper we present  a method to derive interstellar extinction  
using isochrones  from Bertelli et al. (\cite{Bertelli94}) in combination with DENIS CMDs.  
We show that this method   is appropriate for low as well as for highly  
extincted regions.  Finally we  present a map of the extinction in the inner Bulge between  
\mbox{-8$<$$\ell$$<$8 and -1.5$<$b$<$$1.5^{\circ}$}. The finer details  
of the features seen in the map will be discussed in a subsequent paper.

\section{Observations}  
  
The near infrared data were acquired in the framework of the DENIS survey, in a dedicated observation  
of a large Bulge field, simultaneously (Summer 1998) in the three usual DENIS bands, Gunn-I (0.8\,$\mu$m),  
J (1.25\,$\mu$m) and $\rm K_{S}$ (2.15\,$\mu$m). For  the source extraction we used PSF fitting  
optimised for the crowded  fields (Alard et al. in preparation). For the astrometry, the individual  
DENIS frames were cross-correlated with the PMM catalog (USNO-A2.0). The absolute astrometry is then  
fixed by the accuracy of this catalog ($\sim$ 1''). The internal accuracy of the DENIS astrometry,  
derived from the identifications  in the overlaps is of the order of 0.5''.

For the determination of the zero point  all standard stars observed in a given night have been used.  
The typical uncertainty of the zero points has been derived from the overlapping regions and is about  
0.05, 0.15 and 0.15 mag in the I,J and $\rm K_{S}$ bands respectively.  
  
\section{Extinction determination using isochrones}

 Colour-Magnitude Diagrams were constructed for  
sources in a small window (radius of 2 arcmin)  in the field. The modal value of the distribution of the $\rm A_V$ required to  
move the stars in the CMD to the zero extinction isochrone was taken as the value of the extinction for  
this window.  The interstellar extinction law \mbox{($\rm A_{V}:A_{J}:A_{K_{S}} = 1:0.256:0.089$)} from Glass  
(\cite{Glass99}) was used.  
The window was then displaced laterally in uniform steps to construct an image of the  
spatial distribution of the extinction over the whole field.  We have found that towards the  
Galactic Bulge a sampling window of radius 2 arcmin provides a sufficient number of stars to form  
a sequence enabling a reliable estimate of the extinction in such a window.  
We presently do not use the DENIS data with $\rm K_{S} < 7$ due  
to the saturation of the detectors.  
We only use those sources which have been detected in J as well as in $\rm  
K_{S}$ with $\rm K_{S}$ brighter than 11\,mag in order  
to be as complete as possible in J.  This criterion is also quite  
important in order to rule out fake  sources at the fainter end of the luminosity function  
and to guarantee an RGB/AGB identification.  However, in the regions with very high extinction ($\rm  
A_{V} > 25$) we find that a large proportion of the sources detected at $\rm K_{S}$ do not have counterparts at  
the shorter wavelengths. The location of these 'missing' J sources is concentrated in regions  
with high extinction ($\rm A_{V} > 25^{m}$).  
Hence for regions with large extinction, where the number of sources detected  
with the above criterion is smaller than in regions with lower extinction, we only get a lower limit on  
the extinction.  
  
  
 Results with such a sampling  
window and with displacement steps of 1 arcmin are discussed below.  
  
Isochrones from Bertelli et al. (\cite{Bertelli94}) placed at 8\,kpc distance for a 10\,Gyr stellar  
population with Z=0.02 has been used as a reference system. The isochrones were calculated for  
the ESO filter system by convolving the near-infrared bands with the spectral energy distributions from  
Kurucz  (\cite{Kurucz92}) for temperatures higher than 4000\,K. At lower temperatures  
the effective temperature scale from Ridgway et al. (\cite{Ridgway80}) for the  
late M giants and the  Lan\c{c}on \& Rocca-Volmerange (1992) scale for the early M giants has been used.  
The lack of very red standards limits the  near-infrared colour transformation  
(Bressan \& Nasi \cite{Bressan95}) and causes the colours of the Z=0.02  isochrone to 'saturate'  
around $\rm (J-K)_{0}= 1.35$. Therefore a new, empirical $\rm T_{eff}$-(J--K) colour relation  
has been derived by making a fit through the $\rm T_{eff}/(J-K)_{0}$ data available for cool  
giants  [see Schultheis et al. \cite{Schultheis98} and Ng et al. (in preparation)].  
Schultheis et al. (\cite{Schultheis98})  demonstrated the good agreement of the isochrones  
with the new $\rm T_{eff}$-colour relation using NIR photometry of  a sample of Miras and Semiregular  
Variables in a field located at the outer Bulge. However, the upper part above the RGB tip  
($\rm K_{S} \sim 8.0$) remains nevertheless more uncertain.  
  
Based on observed near  
infrared spectra for a sample of M giants and Mira Variables, kindly provided by A.~Lan\c{c}on,  
we have found the difference between K and $\rm K_{S}$ to be small, in the order of  
0.04-0.05\,mag.  
\subsection{Uncertainties of the isochrone method}  
\begin{figure}  
\epsfysize=6.0cm  
\centerline {\epsfbox[25 25 566 753]{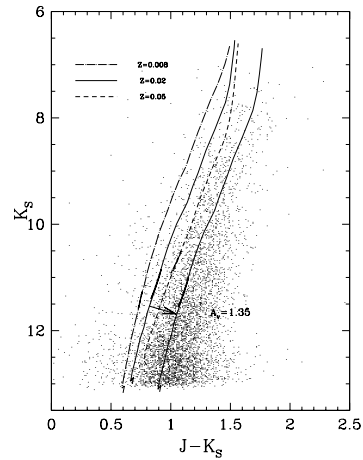}}  
\caption{DENIS $\rm K_{S}/ (J-K)_{S}$ diagram for Baade's window (Sgr I). Superimposed are isochrones with Z=0.008, Z=0.02 and Z=0.05, respectively (see  
text). Saturated stars ($\rm K_{S} < 7$) are not displayed.}  
\label{CMD}  
\end{figure}

Figure \ref{CMD} shows a DENIS CMD in a part of the Baade's Window (SgrI, $\rm \sim 0.06\,deg^{2}$).  Overlaid  
are the isochrones with different metallicities (for $\rm A_{V} = 0$).  The isochrone with  
Z=0.02 is found to follow well the observed CMD and is shown at $\rm A_{V} = 0$ and  
also at $\rm A_{V} = 1.35$ which is the modal value of the extinction found for the  
field.

It is a well known fact that in the Galactic Bulge one deals with a wide  
metallicity and age range.   Using isochrones with a different Z would give us  
different extinction values. In order to estimate  the effect of metallicity on  
the derived extinction values, isochrones with Z=0.05 and Z=0.008 have  been  
 considered.  
Between the isochrone with Z=0.008 and Z=0.05 there is a $\rm  
\Delta(J-K) \sim  0.2$  (see also e.~g.~Bressan et al.~\cite{Bressan98}) which  
corresponds to a $\rm \Delta A_{V} \sim 1^{m}$.  This is the typical  
uncertainty  in the determination of the extinction assuming solar metallicity.  
In contrast to the effect of metallicity the isochrones are hardly affected by an age-range  
(see e.~g. Schultheis et al. 1998).  
\begin{figure*}  
\epsfxsize=5.5cm  
\centerline {\epsfbox[25 154 581 714]{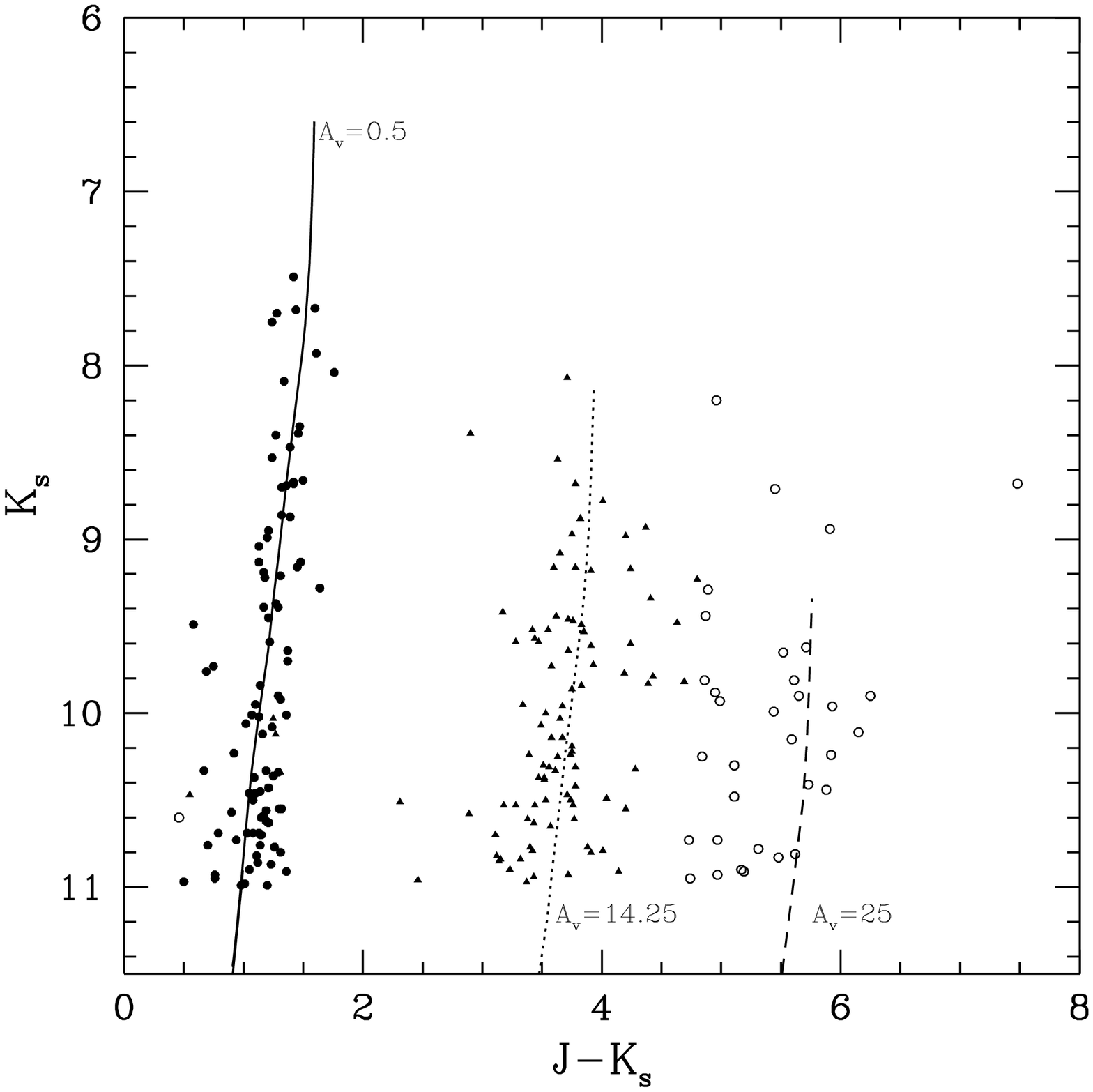} \epsfxsize=5.5cm \epsfbox[25 154 581 714]{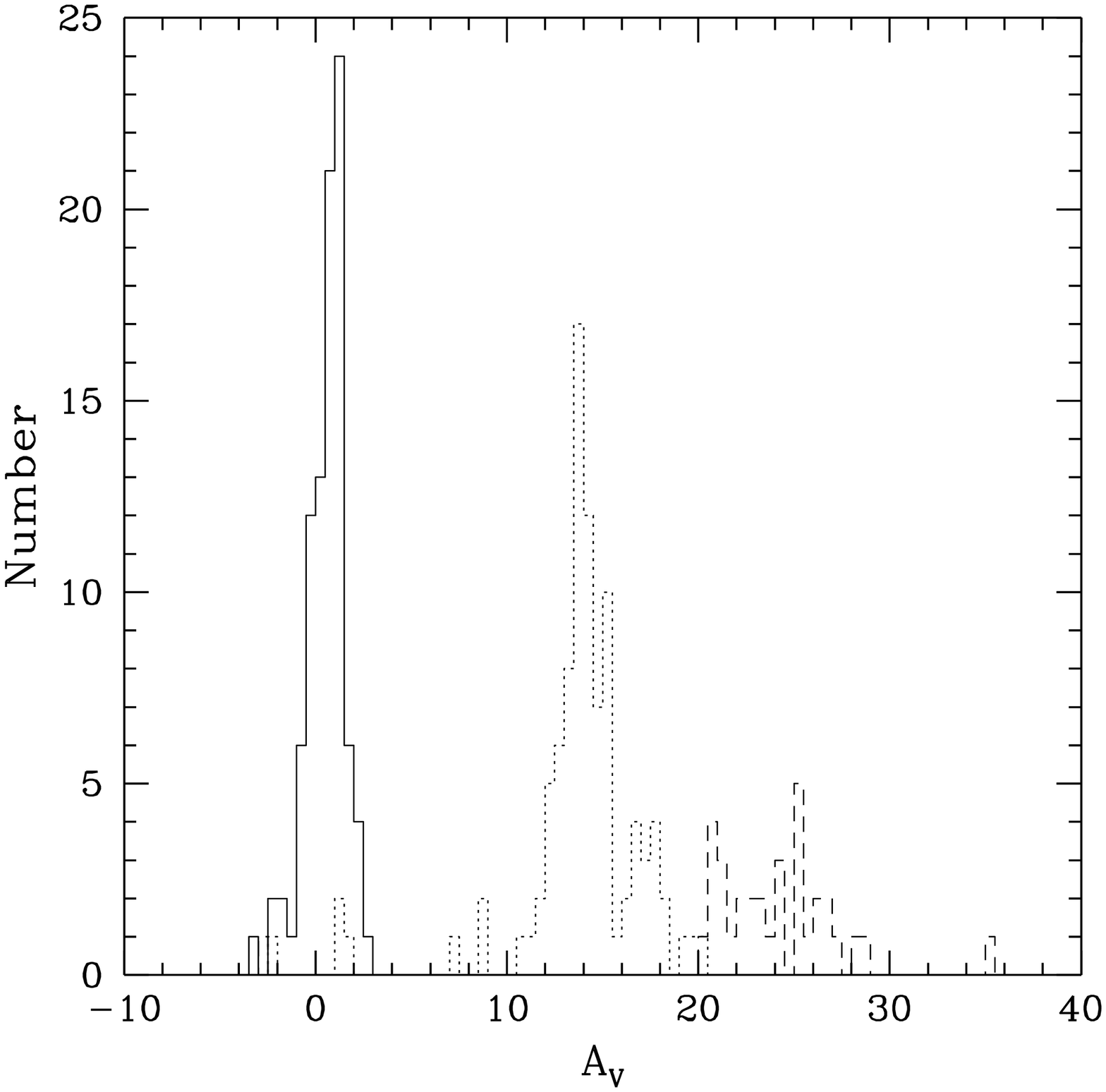}}  
\caption{Observed CMDs with isochrones at derived extinction values (left panel)  
and histograms of the extinction values for individual sources (right panel) for 
sample fields (radius=2') in Baade's window (filled circles, solid line),  
a field located at $\ell$=-0.37 and 
$b = 0.5$ (filled triangles, dotted line) and around the  
Galactic Centre (open circles, dashed line).} 
\label{Figure4} 
\end{figure*}  
  
Analysis of repeated observations (1996 \& 1998) shows that the internal dispersion  
 in the photometry, in the crowded regions, is less than 0.15\,mag for $\rm K_{S} < 11$.  
Compared to the distance spread of the Galactic Bulge ($\rm \sim 0.35\,mag$) the  
errors in the photometric accuracy as well as the errors coming from  
the isochrones are negligible.  
\begin{figure*}  
\resizebox{\hsize}{!}{\includegraphics{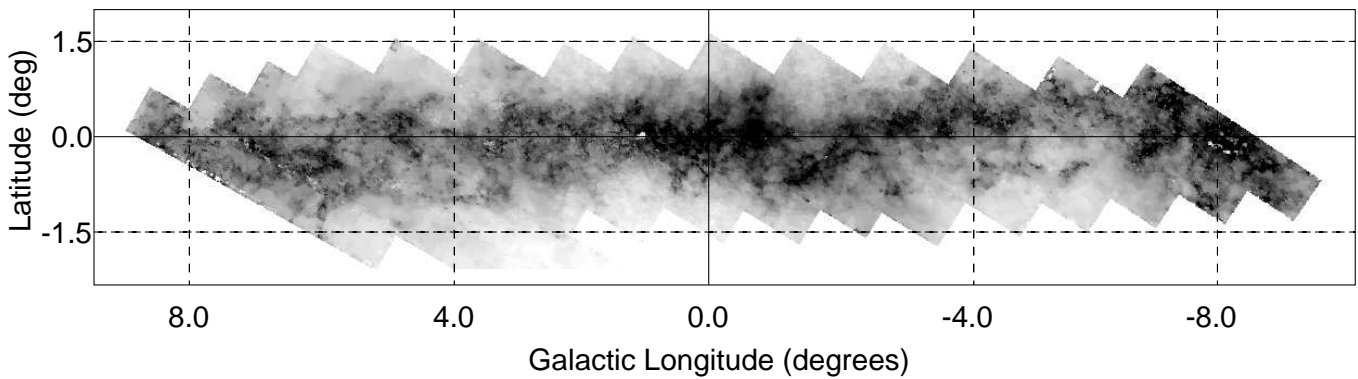}}  
  
\caption{Map of the extinction in the inner Galactic Bulge.  
The image grey scale represents an  
$\rm A_{V}$ range in magnitudes from 0 (white)  to 35 (black). 
A coloured high resolution image can be found at http://www-denis.iap.fr/articles/extinction/} 
\label{Extinction} 
\end{figure*} 
 
\subsection{CMDs in sampling windows} 
 
In Figure \ref{Figure4}  we show the CMDs for sampling windows at three  
different locations, namely a field with low extinction in  Baade's window, one 
with intermediate extinction at $\ell = -0.37^{o}$ and $ b = 0.5^{o}$ and one 
at the Galactic Centre ($\ell = 0, b=0$).    The low extinction field considered here  
is a small part of the Sgr I field whose CMD is presented in Figure  
\ref{CMD}.   While the RGB/AGB branch in Baade's window shows a narrow RGB/AGB 
sequence, for the two other fields one sees clearly a wide spread  in the $\rm 
(J-K_{S})$ colour.  The Baade's window is known to have low and well behaved 
extinction and this is seen as a sharp peak in the distribution of $\rm A_{V}$ 
(figure \ref{Figure4}).  For the field at $\ell = -0.37, b = 0.5$ and around 
the Galactic Centre one does not find a single well defined peak but perhaps 
two different peaks which indicates that there maybe two or more distinct 
layers of extinction causing material along this line of sight. Alternatively 
the absorbing matter may show clumpiness on a scale smaller than the 4 
arcminute diameter of the sampling window. As mentioned earlier, our J and K 
detections are complete in regions with low extinction (up to $\rm A_{V} < 
25$) and a large proportion of K sources do not have J counterparts (up to 
60-90\%) in the regions with high extinction (GC).   It is unfortunately 
not possible to use the results from the regions with low extinction to assign 
a completeness limit for the obscured ones because the controlling factors are 
confusion in the first case and the detector sensitivity in the second.  A map 
of the 'missing' J sources shows that they are a significant source of 
uncertainty in regions with $\rm A_{V} > 25$.

\section{Results}  
\label{results} 

The whole map for the inner Galactic Bulge is shown in Fig.~\ref{Extinction}.  
The map has a resolution of 4 arcminutes (the diameter of the sampling window
used).    Note the clumpy and filamentary behaviour of the distribution of
extinction especially close to the Galactic plane.  There are also small
pockets with uniform extinction of about $\rm A_{V} = 6$ magnitudes (for  
example the ISOGAL field $\ell=0, b=1$ studied by Omont et al. 1999).   Table
{\ref{table}} gives the values of the mean extinction in various locations in
the inner Bulge.  The distribution of extinction along the line of sight
through the Bulge will be discussed in detail in a subsequent paper (Ganesh et
al in preparation).

\begin{table}
\caption{Mean value  of $A_{V}$ (in $0.1^\circ \times 0.1^\circ$ box) along different lines of sight towards the
inner Galaxy}
\begin{center}
\begin{tabular}{rccccccc}
$b~$~\vline&$\ell=6^\circ$&$\ell=3^\circ$&$\ell=0^\circ$&$\ell=-3^\circ$&$\ell=-6^\circ$\\
\hline
$~0.75^\circ$~\vline&6.8&8.9&12.0&11.9&13.7\\
$~0.00^\circ$~\vline&11.7&16.3&25.7&22.8&11.6\\
$-0.75^\circ$~\vline&14.7&10.9&10.9&13.5&7.7\\

\end{tabular}
\end{center}
\noindent
\label{table}
\end{table}

\begin{figure}[h] 
\resizebox{\hsize}{!}{\includegraphics{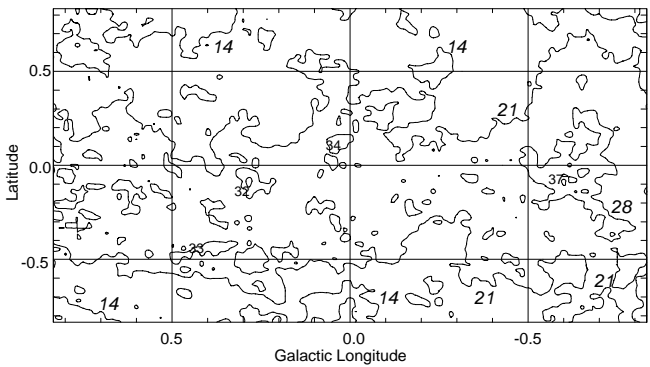}} 
\caption{Contour plot of the extinction around the Galactic Centre with 
contours plotted in the range from 0 to 35 at 7\,mag intervals in 
$\rm A_{V}$.  The major contours are labeled in large italic fonts.   
Local maxima (with $\rm A_{V} > 30$) are also marked with the local $\rm A_{V}$ values in small 
fonts.} 
\label{E_CEN} 
\end{figure}

In Fig.~\ref{E_CEN} we present a contour map of the extinction around the Galactic Centre. 
The distribution of extinction in this region has been studied earlier by Catchpole 
et al (\cite{Catchpole90}) using H\&K data. We find a good agreement in the observed structures 
keeping in mind the higher resolution results presented here. The difference in $\rm A_{V}$ compared 
to Catchpole et al. is typically smaller than 3.  The average interstellar extinction along the Galactic 
plane (see Fig.~\ref{E_CEN}) is $\rm \approx 24^{m}$ in $\rm A_{V}$ although this might 
be a lower limit (as seen by the presence of patches with $\rm A_{V} > 30$) due to the effect of the 'missing' J sources discussed earlier. 
 Along the minor axis the $\rm A_{V}$ decreases to $\rm 15^{m}$ towards the edges, starting from 
 nearly $\rm 25^{m}$ near the Centre. 
The derived extinction values from Fig.~\ref{E_CEN} are also qualitatively in agreement with the values found by Wood et al. (1998) near 
the Galactic Centre.

\section{Conclusion} 
 
DENIS observations in the J and $\rm K_{S}$ band together with isochrones by Bertelli  
et al. (\cite{Bertelli94}) are an excellent tool to map the interstellar extinction. 
A comparison with the field studied by Catchpole shows a very good  
agreement although we see more details due to the better resolution.  
Several fields with relatively low ($\rm A_{V} \sim 6$) and homogeneous extinction can be identified  
on the basis of the extinction map. This identification should help further detailed  
investigation of the stellar population in these windows.  
A study of the three-dimensional distribution of the material responsible for the  
interstellar extinction should be facilitated by the availability of the extinction map.  
  
{\it Acknowledgements}\\ 
{{ 
We want to thank I.~S.~Glass, J.~van Loon and Y.~K.~Ng for useful discussions.  The referee, 
G. P. Tiede, is thanked for useful comments and suggestions.  SG was supported by a fellowship from the Minist\'ere des 
Affaires Etrange\'res, France. MS acknowledges the receipt of an ESA 
fellowship.  
}} 
 {{ 
 The DENIS project is partially funded by European  
Commission through SCIENCE and Human Capital and Mobility plan grants.  
It is also  
supported, in France by the Institut National des Sciences de l'Univers,  
the Education Ministry and the Centre National de la Recherche Scientifique,  
in Germany by the State of Baden-W\"urtemberg, in Spain by the DG1CYT, in Italy  
by  the Consiglio Nazionale delle Ricerche, in Austria by the Fonds zur F\"orderung  
der wissenschaftlichen Forschung und Bundesministerium f\"ur Wissenshaft und  
Forschung, in Brazil by the Foundation for the development of Scientific  
Research of the State of Sao Paulo (FAPESP), and in Hungary by an OTKA grant  
and an ESOC\&EE grant. }}

\end{document}